# Resonance amplification of left-handed transmission at optical frequencies by stimulated emission of radiation in active metamaterials

Zheng-Gao Dong,[1] Hui Liu,[2,*] Tao Li,[2] Zhi-Hong Zhu,[2] Shu-Ming Wang,[2] Jing-Xiao Cao,[2] Shi-Ning Zhu,[2] and X. Zhang[3]

[1]*Physics Department, Southeast University, Nanjing 211189, People's Republic of China*

[2]*National Laboratory of Solid State Microstructures, Nanjing University, Nanjing 210093, People's Republic of China*

[3]*25130 Etcheverry Hall, Nanoscale Science and Engineering Center, University of California, Berkeley, California 94720-1740, USA*

*\*Corresponding author: liuhui@nju.edu.cn*

**Abstract:** We demonstrate that left-handed resonance transmission from metallic metamaterial, composed of periodically arranged double rings, can be extended to visible spectrum by introducing an active medium layer as the substrate. The severe ohmic loss inside metals at optical frequencies is compensated by stimulated emission of radiation in this active system. Due to the resonance amplification mechanism of recently proposed lasing spaser, the left-handed transmission band can be restored up to 610 nm wavelength, in dependence on the gain coefficient of the active layer. Additionally, threshold gains for different scaling levels of the double-ring unit are investigated to evaluate the gain requirement of left-handed transmission restoration at different frequency ranges.

## 1. Introduction

Surface plasmon amplification by stimulated emission of radiation, shortened as spaser [1], was proposed to explore the potential applications in nanoplasmonics. More recently, as an analogy of the localized surface plasmon amplification, a lasing spaser [2], by combining the metallic metamaterial and gain medium, can achieve transmission amplification or lasing effect with coherent stimulated emission. As it is different from the *dark mode* of a spaser, a lasing spaser with metallic metamaterial as the resonant inclusion can radiate to the far field and generate *transmission amplification* with orders of magnitude enhancement at resonance frequency.

On the other hand, metamaterials have attracted wide interest because of their important applications in negative refraction [3-7], superlens [8] and invisible cloak [9]. However, electromagnetic resonance is usually required for the metallic metamaterials to obtain these extraordinary properties [10-13], such as the left-handed transmission, which unavoidably results in severe resonance loss, especially at optical frequencies where metals can not be regarded as perfect conductor. Therefore, how to compensate the intrinsic ohmic loss in metallic metamaterials is very important in order to obtain lossless negative refraction behavior and optical left-handed transmission, because metal attenuation progressively dominates the resonant response at optical frequencies. As a matter of fact, Sarychev and Tartakovsky have demonstrated theoretically that a plasmoic metamaterial comprising metallic horseshoe-shaped nanotennas, if filled with highly efficient gain medium, can serve as a very compact source of electromagnetic radiation [14]. In this paper, inspired by the concept of the lasing spaser mechanism [2] as well as the other works reported earlier, we investigate numerically the amplification of the left-handed resonant transmission at terahertz frequencies, and even visible spectrum, with the aim to compensate the severe intrinsic ohmic loss by stimulated emission of radiation.

## 2. Numerical model

According to the work of Zheludev *et al.*, the active medium, such as a semiconductor pumped by optical or electrical signal, is characterized by a gain coefficient $\alpha$ with the value of $(2\pi/\lambda)\mathrm{Im}(\sqrt{\varepsilon'+i\varepsilon''})$, where $\varepsilon'$ and $\varepsilon''$ are the real and imaginary part of the electric permittivity of the active medium, respectively [2]. In this work, such an active medium is used as the substrate of the metallic structure composed of concentric double rings, which was demonstrated to be efficient on generating the left-handed resonance in microwave regime [15]. Figure 1 shows schematically the unit cell of the double-ring metamaterial, where the geometric parameters *a* and *b* are 3.0 and 0.2 mm, respectively. The gap between the inner and outer rings, represented by *g*, is 0.1 mm. The 0.02-mm-thick rings in square shape are silver with Drude-type dispersion ($\omega_p = 1.37\times10^{16}\,\mathrm{s}^{-1}$ and $\gamma = 8.5\times10^{13}\,\mathrm{s}^{-1}$, see Refs. 16 and 17). As for the active substrate, the thickness is 1.0 mm, the real part of permittivity is 2.1, and the gain coefficient is 450 $\mathrm{cm}^{-1}$. In our numerical simulations based on the full-wave finite element method, the simulation configuration has a dimension of 1×1×5 units (i.e., one unit in the transverse *xy*-plane and five units in the electromagnetic propagation direction along the *z*-axis). The polarized incident wave with electric field in the *x* direction and magnetic field in the *y* direction is satisfied by applying the perfect electric and magnetic boundary conditions, respectively [18-20].

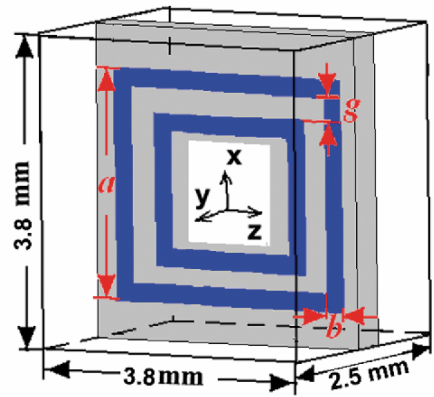

Fig. 1. (Color online) Unit cell of the metallic double-ring metamaterial with an active medium as the substrate. The millimeter sizes correspond to a scaling factor *s*=1.

## 3. Simulation results

As demonstrated in our previous work [15], the double-ring metamaterial with the millimeter sizes specified earlier can exhibit the left-handed behavior with high transmission magnitude. However, by scaling the structure simultaneously down to the micrometer sizes (with a scaling factor $s=10^{-3}$), the strong resonance transmission with left-handed behavior, originally exhibited in gigahertz range (near 26 GHz, see Ref. 15), is not inherited to the terahertz spectrum [see Fig. 2, solid circle line]. This attenuation of resonance transmission is easy to understand because the silver gradually loses its metallic luster at higher frequencies, as described by the Drude-type dispersion. To restore the fading left-handed resonance transmission by compensation of the resonance attenuation, a gain coefficient $\alpha=450\,\mathrm{cm}^{-1}$ is introduced to the active substrate layer. The solid square line in Fig. 2 shows the gain-assisted transmission spectrum, from which it is obvious that the resonance response around 23.8 THz, in consistent with the scaling factor $s=10^{-3}$, restores the transmission profile of its microwave counterpart to a large degree. Notice that there are also transmission amplifications around 18.6 and 33.4 THz, corresponding to the electric resonance frequency $f_{eo}$ and the electric plasma frequency $f_{ep}$ [15,21,22].

The corresponding reflection spectrum is also shown in Fig. 2, where the nonresonant reflectance larger than 0 dB is attributed to the effect of the gain layer. It is interesting that the left-handed transmission amplification at 23.8 THz is occurred with a decrease of the reflectance, while simultaneous enhancements of the transmittance and reflectance are found at both the electric resonance and electric plasma frequencies [see Fig. 2, open square line]. Although Fig. 2 indicates merely that certain gain coefficient is sufficient to compensate the resonance attenuation of metallic double rings above terahertz frequencies, it will be shown in the next part that further magnitude amplification of the resonance transmission is possible by tuning the gain coefficient.

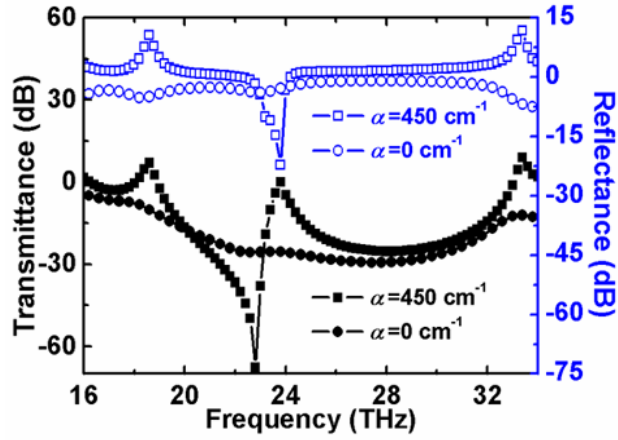

Fig. 2. (Color online) Transmission and reflection spectra of the double-ring metamaterial (scaling factor $s=10^{-3}$) with and without gain coefficient.

For further demonstration of the resonance restoration by using active medium, Figs. 3(a)-3(f) show the magnitude comparisons of the left-handed resonance with and without the gain coefficient. Firstly, the anti-parallel induced current on the metallic surfaces between the inner and outer ring edges implies a magnetic resonance of antisymmetric mode [4,23]. Secondly, after the introduction of gain coefficient $\alpha=450\,\text{cm}^{-1}$, the induced current as well as the resultant localized magnetic field is sufficiently enhanced in magnitude, which leads to the negative-permeability response and thus contributes to the restoration of the left-handed resonance transmission around 23.8 THz. Thirdly, there are two pairs of antisymmetric currents formed by the inner and outer ring, corresponding to magnetic moments of opposite directions, this should reduce, as a whole, the effective magnetic characteristic. However, from the Figs. 3(c) and 3(f), the magnetic alignments are strongly concentrated in the left and right gaps, and it seems that they mainly canceled each other only in the middle part of the double rings. In overall, the simulation result reveals that the opposite magnetic alignments are coupled to form a counterclockwise rotation [see Figs. 3(c) and 3(f)]. Theoretically, such antiparallel magnetic moments in a unit cell could be interpreted in the viewpoint of resonance hybridization, corresponding to the antisymmetric mode $|\omega_+\rangle$ [24,25].

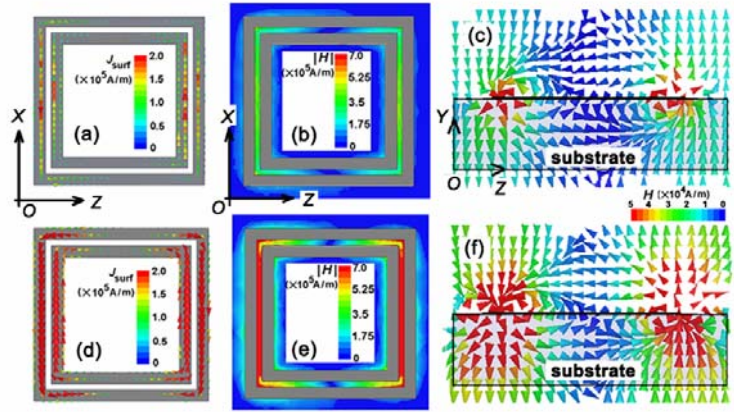

Fig. 3. (Color online) Comparison of the magnetic resonance at 23.8 THz with and without gain coefficient. Induced current, magnitude of magnetic field, and magnetic map at $\alpha = 0 \, \text{cm}^{-1}$ are presented in (a), (b), and (c), respectively. (d), (e), and (f) are the corresponding plots at $\alpha = 450 \, \text{cm}^{-1}$.

Another issue worthy of investigation is the amplitude of resonance restoration and/or amplification as a function of the gain coefficient of the active layer. In fact, the enhancement of the resonance transmission does not always keep increasing with the gain level of the active medium; the reason has been demonstrated in Ref. 2. In our simulations, the peak transmissions of the left-handed resonance around 23.8 THz under different gain coefficients are shown in Fig. 4(a), from which the largest enhancement is obtained at the gain value of $600 \, \text{cm}^{-1}$, corresponding to the maximum resonance amplification of 16.7 dB. Such an amplification magnitude is much smaller than the peak transmission of 42 dB at resonance frequency of about 35 THz from the asymmetric split-ring metamaterial (see Ref. 2). On the other hand, the threshold gain, i.e., the gain coefficient corresponds to 0 dB resonance transmission (so called resonance restoration), about 450 $\text{cm}^{-1}$ at 23.8 THz resonance as shown in Fig. 4(a) is larger than the threshold gain of 70 $\text{cm}^{-1}$ from the asymmetric split-ring metamaterial at 35 THz [2]. The reason for these two aspects may be that the asymmetric split-ring metamaterial has a resonance with higher $Q$-factor than the double-ring one does. In addition, the dependence of the left-handed bandwidth on the gain coefficient can be estimated from Fig. 4(b). The simulation result seems that transmission amplification does not lead to obvious change in the transmission bandwidth, which is attributed to the relatively low *Q*-factor of the double-ring metamaterial.

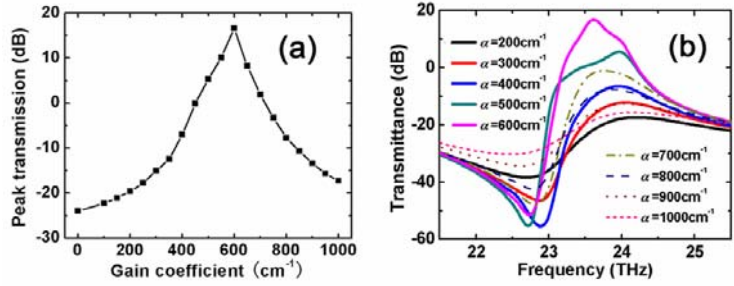

Fig. 4. (Color online) (a) The transmission peak as a function of the gain coefficient at resonance frequency of 23.8 THz. (b) The transmission bands for different gains.

We have demonstrated that the resonance restoration to 0 dB transmission at 23.8 THz corresponds to a threshold gain of 450 $cm^{-1}$. It should be of interest to investigate the threshold gains for different scaling levels of the double-ring unit, in order to evaluate the gain requirement of the left-handed transmission restoration at different frequency ranges, as well as to explore whether the similar left-handed resonance amplification can be generated at the visible spectrum. Figure 5(a) shows the dependence of the threshold gain on the resonance frequency of the left-handed transmission, as a result of shrinking down the sizes of the double-ring unit simultaneously by the scaling factor *s* (in accordance with the scaling law $f_o \propto 1/\sqrt{s^2 + \text{const.}}$, see Ref. 26). In order to obtain the similar resonance behavior in the visible regime, a nanometer version of the double rings is required. However, simply scaling down the microwave metamaterial (see Fig. 1) into the visible spectrum will lead to the unrealistic case of gap *g* and metal thickness smaller than 20 nm. Thus, this case is calculated with a different geometric configuration: edge length of the out ring *a* is 185 nm, gap *g* is 20 nm, metal cross section is 20×20 $nm^2$, and unit cell is in 210×100×210 $nm^3$. Figure 5(b) shows the restoration of the left-handed resonance near 492 THz (about 610 nm wavelength of red light); with the threshold gain

$\alpha = 1.45 \times 10^4$ cm$^{-1}$ and the thickness of the active substrate is 60 nm. It should be emphasized that the proposed gain-assisted left-handed response in visible spectrum is promising for its lossless or even amplification capability of the left-handed resonance transmission; while in contrast, the other visible left-handed metamaterials in literature are confronted with heavily lossy predicament. Nevertheless, It should be reminded that the active media with very high gain coefficients are disadvantageous to be obtained, although the threshold gain for visible spectrum ($\alpha = 1.45 \times 10^4$ cm$^{-1}$) is comparable to that of the active media available in semiconducting quantum dots. For example, gain with order of $10^5$ cm$^{-1}$ has been attainable experimentally [27,28]. Therefore, increasing the *Q*-factor of the left-handed resonance by exploring optimization of metallic metamaterials, consequently decreasing the threshold gain, should be important for improvement of such gain assisted left-handed resonance transmission.

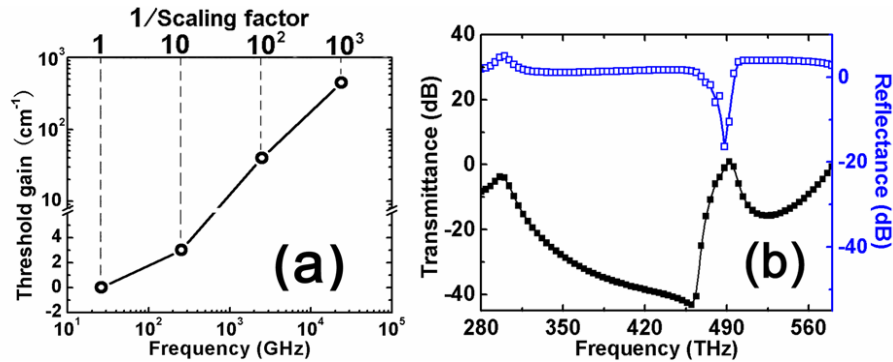

Fig. 5. (Color online) (a) The threshold gain as a function of resonance frequency, corresponding to different scaling factors. (b) Left-handed resonance transmission at 492 THz, with the threshold gain $\alpha = 1.45 \times 10^4$ cm$^{-1}$.

## 4. Conclusions

Resonance amplification of left-handed transmission with considerable magnitude enhancement is demonstrated numerically in a metamaterial composed of metallic double rings when an active medium is introduced as the substrate. The significant amplification of left-handed transmission remains very well even when the left-hand resonance moves up to the visible spectrum, indicating that it is possible to overcome the severe resonance loss inside metals at optical frequencies. By optimizing the metallic structures of various left-handed metamaterials, left-handed transmission amplification with less threshold gain value could be obtained in the future. Amplification method similar to the active left-handed materials could also be extended to other metamaterials, such as superlens and cloaking structures.

**Acknowledgments**

This work was supported by the State Key Program for Basic Research of China (Nos. 2004CB619003, 2006CB921804 and 2009CB930501) and the National Natural Science Foundation of China (Nos. 10534020, 10604029, 10704036, 10747116, 60578034 and 10874081).